\documentclass{article}
\usepackage{spconf,amsmath,graphicx}
\usepackage{gensymb}

\def\L{{\cal L}}

\title{Learning to Estimate Kernel Scale and Orientation of Defocus Blur with Asymmetric Coded Aperture}
\name{Jisheng Li $^*$$^{1}$\thanks{$^*$ Denotes equal contribution, and $^{\dag}$ denotes corresponding author.}, Qi Dai $^*$$^{3}$, Jiangtao Wen$^{2,1,\dag}$}
\address{$^{1}$Tsinghua University \\
$^{2}$Research Institute of Tsinghua University in Shenzhen, Shenzhen, China \\
    $^{3}$Boyan Technology DBA RayShaper China}

\begin{document}
\ninept
\maketitle
\begin{abstract}
Consistent in-focus input imagery is an essential precondition for machine vision systems to perceive the dynamic environment. A defocus blur severely degrades the performance of vision systems. To tackle this problem, we propose a deep-learning-based framework estimating the kernel scale and orientation of the defocus blur to adjust lens focus rapidly. Our pipeline utilizes 3D ConvNet for a variable number of input hypotheses to select the optimal slice from the input stack. We use random shuffle and Gumbel-softmax to improve network performance. We also propose to generate synthetic defocused images with various asymmetric coded apertures to facilitate training. Experiments are conducted to demonstrate the effectiveness of our framework. 
\end{abstract}
\begin{keywords}
Blur kernel estimation, focus adjustment, asymmetric coded aperture
\end{keywords}
\section{Introduction}
\label{sec:intro}

Defocus blur is common in the digital image capture process but unwelcome in the machine vision system, as it degrades features and leads to a system accuracy drop. The imaging techniques and optical performance have improved rapidly during the past decade, but the defocus problem remains to be solved. Consumer cameras handle this problem by introducing autofocus functions. In contrast, cameras like surveillance cameras lack autofocus function or have a fixed focus distance.
While seeing remarkable results on curated web image datasets like ImageNet~\cite{deng2009imagenet}, a real-world application often fails to achieve a comparable accuracy using footage frequently contains out-of-focused imagery.
In this work, we took a step further toward real-world application and proposed an end-to-end framework to estimate the defocus blur scale and orientation for auto lens focus adjustment.

Conventional methods to solve out-of-focus problems can be categorized into two approaches:  introducing a focus tracking mechanism~\cite{he2003modified} or performing a deblurring algorithm~\cite{levin2007image}\cite{lucy1974iterative} to recover the all-in-focus image. 
A focus tracking mechanism enables the vision system to capture a focused image directly, in which the optimal location of the focal plane should be determined in advance in order to adjust the lens focus. The focal plane location can be determined either by estimating the blur kernel~\cite{levin2007image} or object distance~\cite{dorsel2002apparatus} (estimation method) or through repetitive tuning~\cite{he2003modified} (trial and error method), i.e., dialing the focal plane back and forth. We notice that a lens adjustment could perform accurately without iteration yielding lower delay using the estimation method. In comparison, the trial and error method takes a few frames to converge, making the estimation method an ideal option. 

In computational photography literature, coded aperture techniques are used for both defocus deblurring~\cite{zhou2009coded}\cite{zhou2009good} and depth estimation~\cite{levin2007image}\cite{sellent2014side}. Though the deblurring functionality seems to solve the defocus problem straightforwardly, an incorrect point spread function(PSF) used in the deconvolution can cause more artifacts against our original intention. Meanwhile, whether deblurred images can deliver the same precision on machine vision tasks as in-focus images do stay unknown. The other practice of coded apertures, depth estimation, can well serve the blur scale estimation in our framework, though very few efforts have been put to distinguish which side of the object rests with respect to the focal plane, which causes the \textit{focal plane ambiguity} in real-world applications. 

Inspired by previous works~\cite{zhou2009good}\cite{sellent2014side}, we tackle the defocus blur problem by estimating blur scale and orientation to guide the lens focus mechanism. With a predicted defocus blur scale and orientation, the optical system can respond with rapid lens adjustment, yielding lower delay and better focus accuracy. 
To achieve this, we propose a novel end-to-end framework with a 3D ConvNet to predict the defocus blur scale and orientation at the same time. We use asymmetric coded apertures to better solve the \textit{focal plane ambiguity} problem neglected in previous research. Our proposed pipeline computes a stack of deblurred images with varied PSF hypotheses as input and predicts each pixel's correct blur 
label (i.e., blur size and orientation). With a 3D ConvNet design, our method is applicable to different numbers of blur scale and orientation proposals. A full discussion of such design is conducted in Section~\ref{sec:methodology}. 
The contribution of our work is three-fold: First, we propose a novel deep-learning-based pipeline to construct a deblurred defocused image volume and estimate the scale and orientation from it. Second, we present a defocused image generation method to address the data-hungry problem. Third, we conduct extensive experiments to examine the robustness of this design among different conditions.
\let\thefootnote\relax\footnotetext{We acknowledge funding from the Shenzhen Science and Technology Program(Grant No.KQTD20190929172829742).}
\vspace{-9pt}
\section{Related Work}
\label{sec:format}
Substantial research has been laid out on blur parameter estimations. In this section, we discuss the methodology foundation, as well as its limitations. 

A defocus blur is caused by object distances away from the camera focal plane. In order to establish a one-to-one correspondence between the object distance and the defocus blur, we must solve the \textit{focal plane ambiguity}, i.e., distinguish which side of the object rests with respect to the focal plane. This ambiguity is non-trivial since the commonly used circular aperture employed in cameras will produce identical blur for items resting before and behind the focal plane. In~\cite{sellent2014side}, Sellen~\etal~detailed how asymmetric coded apertures can solve this ambiguity by creating different blur for items at each side of the focal plane. Following this scheme, we adopted an asymmetric coded aperture in our system design, though our pipeline is not rigged with one particular aperture. 

Estimating the object distance or blur scale is a sub-topic of the well-studied coded aperture dilemma. Numerous coded apertures~\cite{levin2007image}\cite{zhou2009good}\cite{zhou2009coded}\cite{sellent2014side}\cite{grosse2010coded}\cite{nagahara2010programmable}\cite{lin2014separable}\cite{bando2008extracting} have been proposed for different intentions. As~\cite{sellent2014side} points out, scenarios that object located at both sides of the focal plane have not been discussed in previous studies. Moreover, these methods are paired with a particular aperture design, which means variant on PSF caused by lens distortion will affect its performance.
With a predicted PSF, defocus deblurring can be performed in either frequency domain~\cite{richardson1972bayesian}\cite{lucy1974iterative} or spatial domain~\cite{levin2007image}\cite{levin2009understanding}. Although the frequency domain process is more computation friendly, methods performed in the spatial domain like the algorithm in~\cite{levin2007image} produce better perceptual results. 

In this work, we designed our pipeline to avoid these limitations. Our framework generates a deblurred image stack from the defocused image captured using one of above mentioned asymmetric coded apertures. As no extra modification is needed when altering coded apertures or deblurring algorithms in the proposed method, we decoupled the performance dependency from them.  We showed that our system could achieve similar performance on different asymmetric coded apertures with two separate deblurring algorithms in Section~\ref{sec:experiment}.

The topic Depth from Defocus (DFD)~\cite{chaudhuri2012depth} is similar to blur scale estimation as well as recent research on defocus map estimation~\cite{zhuo2011defocus}\cite{lee2019deep} or defocus blur detection~\cite{shi2015just}\cite{ma2020defocus}, which also involves the consideration of blur scale estimation. As these topics themselves are broad studies and neglect to address the \textit{focal plane ambiguity}, we did not compare them in this discussion.
\vspace{-9pt}
\section{Methodology}
\label{sec:methodology}

\subsection{Problem Formulation}
\label{formulation}
The input of our pipeline is a \textit{spatially-varying} defocused image. Spatially-varying indicates the blur at each pixel may differ from the neighbor in terms of scale and orientation. It has been shown in ~\cite{masoudifar2016image} that performing a deblurring algorithm with the wrong kernel yields low-quality images while the correct one produces high-quality results. Using this insight, we formulate blur kernel estimation to separate a correct deblurred result from the wrong ones to teach the neural network how to choose the right kernel scale and orientation. Thus, in our framework, we first perform a deblurring algorithm with PSF hypotheses (varying in scale and orientation) on a defocused image. These results are then concatenated and fed into the network, which is trained to identify the optimal deblurred result index for each pixel. The PSF kernel associated with the optimal deblurred image quality reports the correct scale and orientation.

Our training target can be formulated as an optimization problem:
\vspace{-1mm}
\begin{align} \label{equ: optimization}
    \min_\theta \L_c(& pred_\theta, RS(gt)) + \lambda_s \L_s(GS(pred_\theta)), \\
    & pred_\theta = f_\theta(RS(d_1, d_2, \dots, d_n, dI)) \nonumber,
\end{align}
where a weighted sum of cross-entropy loss $\L_c$ and smoothness loss $\L_s$ is minimized. The network (denoted as $f_\theta(\cdot)$, with model parameters $\theta$) takes the deblurred results $d_1, d_2, \dots, d_n$ and the original defocused image $dI$ as input. We perform random shuffle (denoted as $RS(\cdot)$) on the input and the ground-truth label $gt$ to avoid overfitting and impose a smoothness loss on the network output $pred_\theta$ for better performance. In computing the smoothness loss, we apply the Gumbel-Softmax~\cite{jang2016categorical} technique (denoted as $GS(\cdot)$) in order to convert the 3D output $pred_\theta$ into 2D. The smoothness loss is then weighted by $\lambda_s$. We detailed the motivation to perform random shuffle and the Gumbel-softmax methods in Section~\ref{sec: kernel_est}.
\vspace{-9pt}
\subsection{Blur Kernel Estimation}
\label{sec: kernel_est}

\noindent {\bf Input and Output}\, 
With a defocused image as input, we first perform a deblurring algorithm with several PSF  hypotheses to form the deblurred image stack. We adopt the deblurring approach proposed in~\cite{levin2007image}, where the image is deblurred in the spatial domain using the Conjugate Gradient method, as well as the method in the frequency domain using Wiener Deconvolution. There is no extra modification that needs to be made in our framework when changing the deblurring algorithm. 
The deblurred image stack is a concatenation along \textit{depth}  channel of  $d_1, d_2,$ $ \dots, d_n$ and $dI$ with dimension [\textit{height, width, depth, channel}]. 
We further append the stack with $23 - n$ black padding (pixel values with all 0) along the depth dimension, making the depth always 24 for consistent behavior with variable deblurring hypotheses $n$. 
The defocused image $dI$ is also included in the input, as some pixels presented in $dI$ may already be of optimal quality, and there will be no better options in the deblurred image stack. 

The network output is a logit volume with a size of [\textit{height, width, depth}]. Vectors along the depth dimension reveal the likelihood that slices in the input are \textit{sharp} (high-quality)  or not. The final blur scale and orientation for each pixel can be determined by looking up the blur kernel parameter associated with the highest logit value. 

\noindent {\bf Architecture}\, We design a fully convolutional network followed the architecture in~\cite{mildenhall2019llff}, which utilizes the 3D ConvNet to handle the input with a variable length of depth. Our network is trained to award the image with optimal quality in the input stack while punishing lower-quality ones. A 3D ConvNet design suits this purpose perfectly, as sliding the convolution kernel along the depth dimension imitates the behavior of examining the image one by one. In our scheme, each 3D convolution layer is followed by one layer-normalization layer~\cite{ba2016layer}. The architecture is detailed in Table.~\ref{tab:net_arch}.

\begin{table}[!t]
\begin{center}
\resizebox{0.45\textwidth}{!}{
\begin{tabular}{cccccccc}
\thickhline
\textbf{Layer} & \textbf{s} & \textbf{d} & \textbf{n} & \textbf{depth} & \textbf{in} & \textbf{out} & \textbf{input} \\\hline
conv1\_1 & 1 & 1 & 8  & 24/24 & 1 & 1 & stack           \\
conv1\_2 & 2 & 1 & 16 & 24/12 & 1 & 2 & conv1\_1         \\
conv2\_1 & 1 & 1 & 16 & 12/12 & 2 & 2 & conv1\_2         \\
conv2\_2 & 2 & 1 & 32 & 12/6  & 2 & 4 & conv2\_1         \\
conv3\_1 & 1 & 1 & 32 & 6/6   & 4 & 4 & conv2\_2         \\
conv3\_2 & 1 & 1 & 32 & 6/6   & 4 & 4 & conv3\_1         \\
conv3\_3 & 2 & 1 & 64 & 6/3   & 4 & 8 & conv3\_2         \\
conv4\_1 & 1 & 2 & 64 & 3/3   & 8 & 8 & conv3\_3         \\
conv4\_2 & 1 & 2 & 64 & 3/3   & 8 & 8 & conv4\_1         \\
conv4\_3 & 1 & 2 & 64 & 3/3   & 8 & 8 & conv4\_2         \\\hline
nnup\_5  &   &   &    & 3/6   & 8 & 4 & conv3\_3+conv4\_3 \\
conv5\_1 & 1 & 1 & 32 & 6/6   & 4 & 4 & nnup\_5          \\
conv5\_2 & 1 & 1 & 32 & 6/6   & 4 & 4 & conv5\_1         \\  
conv5\_3 & 1 & 1 & 32 & 6/6   & 4 & 4 & conv5\_2         \\  
nnup\_6  &   &   &    & 6/12  & 4 & 2 & conv2\_2+conv5\_3 \\
conv6\_1 & 1 & 1 & 16 & 12/12 & 2 & 2 & nnup\_6          \\
conv6\_2 & 1 & 1 & 16 & 12/12 & 2 & 2 & conv6\_1         \\  
nnup\_7  &   &   &    & 12/24 & 2 & 1 & conv1\_2+conv6\_2 \\
conv7\_1 & 1 & 1 & 8  & 24/24 & 1 & 1 & nnup\_7          \\
conv7\_2 & 1 & 1 & 8  & 24/24 & 1 & 1 & conv7\_1         \\  
conv7\_3 & 1 & 1 & 1  & 24/24 & 1 & 1 & conv7\_2         \\\thickhline
\end{tabular}}
\end{center}
\vspace{-4mm}
\caption{Our proposed network architecture. All convolution kernels have a size of 3. Column \textbf{s} is the stride, \textbf{d} the kernel dilation, \textbf{n} the number of filters, \textbf{depth} the number of layer input/output depth, \textbf{in} and \textbf{out} the accumulated stride of layer input/output, \textbf{input} the layer input with + meaning concatenation along the channel dimension. \textbf{Layers} starting with “nnup” perform 2x nearest neighbor upsampling.}
\vspace{-4mm}
\label{tab:net_arch}
\end{table}

\noindent {\bf Data Augmentation}\, We expect the network to identify the blur kernel by examining input images' quality. However, preliminary experiments show overfitting trends when changing the input stack order, which means the network exploits the ordering pattern but not the image quality we intended. We perform a random shuffle on the input stack and the label map along the depth dimension during training in improved schemes, and the overfitting issue is resolved.

\noindent {\bf Loss}\, Our network training is formulated as a classification problem. We index each proposed blur kernel employed in the deblurring step and compute the predicted softmax from the output logit volume. The cross-entropy $\L_c$ between the label and the prediction can be computed and serve as our primary loss.

We expect the estimated blur map to be smooth as the blur scale and orientation share the same continuous character as scene depth does in most areas. To compute smoothness loss on our output, we need to convert the 3D logit volume into a 2D indexing map. As the $\arg\max$ operation to flat a 3D volume into 2D is not feasible here with its non-differentiable property, we apply Gumbel-Softmax~\cite{jang2016categorical} to fulfill this need. It takes a logit vector as input, and outputs the index of the element associated with the largest logit value. Then, we compute $\L_s$ through taking the absolute average of the first-order gradient of the 2D indexing map. The final loss $\L$ is a weighted sum of $\L_c$ and $\L_s$: $\L = \L_c + \lambda_s \L_s$, where we weight $\L_s$ by $\lambda_s$. 
\vspace{-5pt}
\section{Experiments}
\label{sec:experiment}

\subsection{Data Generation}
Our training procedure needs a large portion of spatially-varying defocused images, each with a per-pixel ground-truth blur scale and orientation to supervise the network training. With very few previous works sharing coded aperture images, none of them involves scene objects resting on both sides of the focal plane. As we intend to address the focal plane ambiguity problem, we need to generate a new dataset to meet the need. Here we present an approach to acquire defocused images from CG generated all-in-focus images. 

\begin{figure}[t]
    \centering
    \centerline{\includegraphics[width=8.5cm, height=4.2cm]{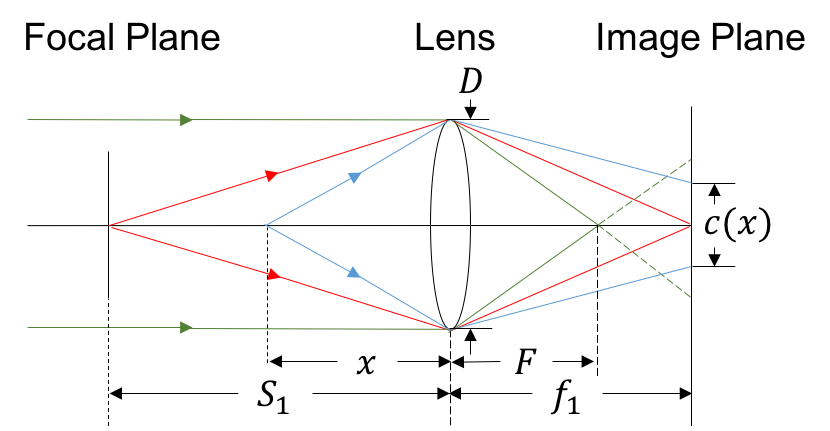}}
    \caption{The thin-lens Model. $S_1$ is the object-space focal distance, $f_1$ the image-space focal distance, $F$ the focal length, $D$ the diameter of aperture. $c(x)$ is the COC diameter for object with distance $x$.}
    \label{fig:thin_lens}
\end{figure}

\noindent {\bf All-in-Focus Image and Defocus Generation}\, We generate all-in-focus images each with its depth map using UnrealCV \cite{qiu2017unrealcv}. Defocused images are further generated following the thin-lens model (Fig.~\ref{fig:thin_lens}) from all-in-focus images, similar to one described in \cite{lee2019deep}. With setting the focal plane at distance $S_1$, the circle of confusion (COC) diameter $c(x)$ for image pixel with depth $x$ can be determined according to Eq.~\ref{equ:coc}.

\vspace{-2mm}
\begin{equation}
    \centering
    c(x) = \alpha\frac{x-S_1}{x},\,\,with\,\,\alpha=\frac{f_1}{S_1}D
    \label{equ:coc}
\end{equation}
\vspace{-2mm}

The sign and magnitude of $c(x)$ indicate the orientation and the scale of the blur kernel, respectively. 
 With an all-in-focus image ${I}_{k}$ and coresponding depth map ${d}_{k}$, for each pixel ${p}_{i} \in {I}_{k}$ and its depth value ${x}_{{p}_{i}}$, we can compute the blur parameter $c({x}_{{p}_{i}})$ according to Eq.~\ref{equ:coc}. We round $\{c({x}_{{p}_{i}})|\forall {p}_{i}\in{I}_{k}\}$ to obtain a set of $\{{c}_{{I}_{k}}\}$ where $[c({x}_{{p}_{i}})]\in\{{c}_{{I}_{k}}\}, \forall {p}_{i}\in{I}_{k}$. Here $[\cdot]$ denotes the rounding operation.
With ${c}_{i}\in \{{c}_{{I}_{k}}\}$ , we resize, flip, and normalize the coded aperture to form a PSF ${F}_{{c}_{i}}$ according to ${c}_{i}$. Convolution is then performed on each pixel ${p}_{i}$ in the all-in-focus image ${I}_{k}$ with kernel ${F}_{[c({x}_{{p}_{i}})]}$ to form the spatially-varying defocused image $\hat{{I}}_{k}$. In addition, we save $[c({x_{{p}_{i}}})], \forall {p}_{i}\in{I}_{k}$ as the per-pixel ground-truth label for the training process. To avoid producing unrealistic blurry images, we restrict the maximum blur size during defocus generation. 

We capture 100 all-in-focus images from scenes provided in UnrealCV, and generate defocused images for each one with 12 different focal distances ($S_1$ in Fig.~\ref{fig:thin_lens}). Thus 1200 defocused images are acquired totally with each coded aperture we used. The maximum blur scale is restricted to be 7, 9, and 10 pixels to form the depth-13, depth-17, depth-19 dataset, respectively. We later use these datasets vary in coded apertures and maximum blur scales to verify our framework's generalization performance. 

\begin{figure}[t]
    \centering
    \begin{minipage}[b]{.48\linewidth}
      \centering
      \centerline{\includegraphics[width=2.0cm]{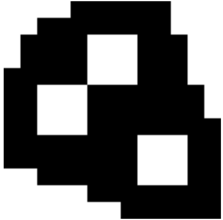}}
      \centerline{(a) Coded aperture from~\cite{sellent2014side}}\medskip
    \end{minipage}
    \hfill
    \begin{minipage}[b]{0.48\linewidth}
      \centering
      \centerline{\includegraphics[width=2.0cm]{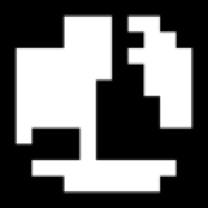}}
      \centerline{(b) Coded aperture~\cite{zhou2009good}}\medskip
    \end{minipage}
    \vspace{-4mm}
    \caption{Asymmetric coded apertures used in experiments}
    \vspace{-4mm}
    \label{fig:coded}
\end{figure}
\begin{figure}[t]
    \centering
    \centerline{\includegraphics[width=8.5cm, height=5cm]{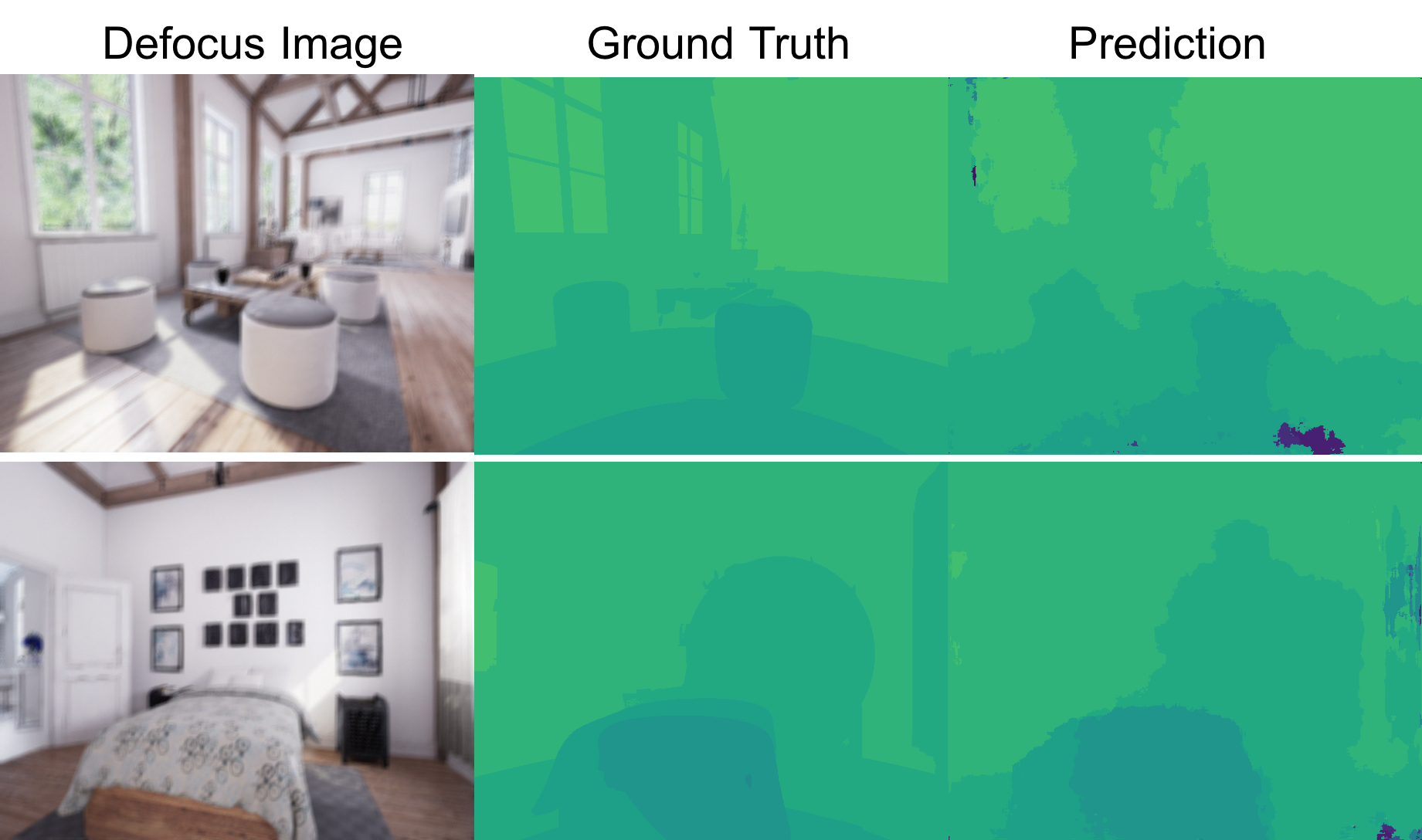}}
     \vspace{-2mm}
    \caption{Qualitative results of two scene picked from our dataset. The middle column is the groud-truth label while the right column is predicted by our framework}
    \label{fig:baseline}
    \vspace{-2mm}
\end{figure}



In our experiments, we split the entire dataset for each coded aperture and maximum blur scale to 936 defocused images for training, 108 for validation, and 156 for evaluation. The split is implemented in such a manner that there are no overlapping scenes across the training/validation/evaluation subsets, which means all 12 defocused images generated from one particular scene are forced to belong to one of these three subsets. 
\vspace{-9pt}
\subsection{Implementiation Details} 

We implement our proposed network architecture in Table.~\ref{tab:net_arch} using TensorFlow API~\cite{abadi2016tensorflow}. In the training procedure, the network is optimized by the SGD optimizer with the random shuffle applied on the input stack and label. The learning rate, batch size, and smoothness weight $\lambda_s$ are 0.01, 1, and 0.1, respectively. The temperature argument for the Gumbel-softmax function is 0.5. We train the network on a single GPU (GeForce RTX 2080 Ti), and the network takes around five days to converge.

\subsection{Evaluation}
\label{sec:evaluation}
We formulate our blur kernel estimation as a classification problem. Logit volume from the network output indicates whether pixels in the deblurred stack are classified as optimal-quality or not. 
The predicted blur scale and orientation can be determined by indexing the highest value along the depth channel of the logit volume.
 We report the prediction N-1 accuracy and the N-3 accuracy in the following manner: N-1 accuracy indicates the percentage of pixels holds the same predicted blur scale and orientation as the ground-truth label does. The N-3 accuracy is computed similar to N-1 but also counting the nearby prediction (i.e., the pixel is predicted within the range of its ground-truth label $\pm 1$) as a correct prediction. For example, pixel with label +3 predicted as +2 (or +4) is not counted as a correct prediction in N-1 but in N-3. 
The intention behind reporting the N-3 accuracy is to compensate for the inaccuracy introduced during rounding COC diameter $c(x)$. The N-3 accuracy also registers how effective this framework will perform in a real-world application. Since even the prediction is off by 1 pixel, adjusting focus with it will still result in a much more in-focus image. 

\begin{table}[t]
\begin{center}
\begin{tabular}{|l|c|c|c|}
\hline
& depth-13 & depth-17 & depth-19 \\
\hline
N-1 Accuracy & 63.08\% & 63.47\% & 63.81\% \\
\hline
N-3 Accuracy & 82.90\% & 82.93\% & 85.59\% \\
\hline
\end{tabular}
\end{center}
\vspace{-4mm}
\caption{Evaluation on input with different valid hypotheses. The model is trained on depth-17, and is evaluated on depth-13, 17 and 19.}
\label{tab:varied_depth}
\end{table}

\begin{table}[t]
\begin{center}
\begin{tabular}{|l|c|c|c|}
\hline
& SoS & SoZ & ZoZ (fine-tuned) \\
\hline
N-1 Accuracy & 63.47\% & 54.16\% & 59.17\% \\
\hline
N-3 Accuracy & 82.93\% & 79.00\% & 81.47\% \\
\hline
\end{tabular}
\end{center}
\vspace{-4mm}
\caption{Evaluation results on dataset generated with different asymmetric coded apertures. SoS means network is trained and evaluated on coded aperture proposed by Sellent \etal~\cite{sellent2014side} . SoZ means trained on Sellent's but evaluated on Zhou's coded aperture\cite{zhou2009good}. ZoZ (fine-tuned) shows the results where the model is pre-trained on Sellent's then fine-tuned and evaluated on Zhou's coded aperture.}
\label{tab:ablation_study}
\vspace{-4mm}
\end{table}
\noindent{\textbf{Evaluation on different number of blur hypotheses}}
As shown in Table.~\ref{tab:varied_depth}, we demonstrate our framework's performance to handle the input with a variable number of blur hypotheses. 
Our model is trained on dataset depth-17 and reports an N-3 accuracy of 82.9\%, 82.93\%, and 85.59\% on evaluation subsets of datasets depth-13, 17, and 19, respectively. The similar high percentage results among three datasets show our framework is able to generalize over input with a variable number of blur hypotheses and estimate the blur scale and orientation of defocused image with high accuracy. 
To challenge our network's generalization capability even more, we alter the coded aperture used in the dataset generation and deblurring step. 
In particular, we train our model on the dataset generated using Sellent's coded aperture and evaluate on two separate datasets generated using Sellent's and Zhou's coded aperture (Fig.~\ref{fig:coded} (a) and (b)). In addition, Sellent's coded aperture is designed for depth estimation, while Zhou's is intended for defocus deblurring. The evaluation accuracy is presented in Table.~\ref{tab:ablation_study}. 
Although we notice the results of SoZ are lower than SoS, dropping from 82.93\% to 79\% in the N-3 accuracy, we observe improved accuracy in ZoZ after the network fine-tuned on the dataset corresponding to Zhou's coded aperture. The performance gap between SoS and ZoZ indicates the coded aperture chosen in the framework would affect the overall performance. Nevertheless, we do not recognize a strong dependency of robust system performance on a particular coded aperture, thus proving our network's generation capability.

\noindent{\textbf{Evaluation on different coded apertures}}
We also assess the network performance with two deblurring algorithms generating the input stack from defocused images. These two algorithms are the spatial-domain method proposed in~\cite{levin2007image} and the Wiener Deconvolution, which is a frequency-domain method. While the model is trained with~\cite{levin2007image}  generating the deblurred image stack, experiments show that swapping to the Wiener Deconvolution algorithm does not influence the performance: The N-1 accuracy of evaluating on spatial-domain vs. frequency-domain deblurring algorithm is 63.47\% vs. 62.48\%, and N-3 accuracy of 82.90\% vs. 82.64\%, showing that our framework is insensitive to the deblurring algorithm type. 

\noindent{\textbf{Qualitative evaluation}}
Some qualitative results are exhibited in Fig.~\ref{fig:baseline}. We plot the blur map similar to a depth map that pixels with similar colors indicate a similar blur parameter. Our network can generally discriminate areas with different blur kernels and classify them accurately. Some pixels are misclassified to a kernel closer to the true one. This is because deblurring with a kernel similar to ground-truth yields a sub-optimal image quality but can be hardly distinguished by the network. The network can not always discriminate between an optimal result (deblurred by the true kernel) and a sub-optimal result. Nevertheless, we achieve 82.93\% N-3 accuracy. This indicates that our network can identify the optimal results and assign a high probability to them. 

After further analyzing the incorrect prediction, which significantly differs from its neighbor region's prediction and the ground-truth label, such as the bottom right part of the first row (dark area) in Fig.~\ref{fig:baseline}, we find that a large piece of them comes from the texture-less region in the all-in-focus image. We hold this is an ill-posed challenge as defocus texture-less images with various blur scale and orientation show a tiny variation in between. Also, in real-world applications, a texture-less region can hardly be a region of interest.
\vspace{-9pt}
\section{Conclusion}
\label{sec:conclusion}
In this paper, we present a deep-learning-based framework to estimate blur scale and orientation from a defocused image to assist rapid focus adjustment. Benefiting from the 3D ConvNet architecture, our pipeline is able to process a variable depth of deblurred candidates across complex scenarios. We employ a random shuffle technique to avoid overfitting and adopted the Gumbel-softmax method with smoothness constraint to improve the performance. Experiments show that our approach achieves reliable accuracy per-pixel blur scale and orientation results. When examined on different coded aperture datasets, our framework generalizes well to achieve above 80\% N-3 accuracy after fine-tuning. Furthermore, our system performance does not depend on a particular deblurring algorithm or coded aperture.
\bibliographystyle{IEEEbib}
\bibliography{main}

\end{document}